# Method for Quantitative Estimation of the Risk Propagation Threshold in Electric Power CPS Based on Seepage Probability

**ZHAOYANG QU[1,2], YU ZHANG[1], NAN QU[3], LEI WANG[1], YANG LI[4,5] (Senior Member, IEEE)，AND YUNCHANG DONG[1]**

[1]College of Information Engineering, Northeast Electric Power University, Jilin 132012, China
[2]Jilin Engineering Technology Research Center of Intelligent Electric Power Big Data Processing, Jilin 132012, China
[3]Maintenaue Company of Jiangsu Power Company, Nanjing 210000, China
[4]School of Electrical Engineering, Northeast Electric Power University, Jilin 132012, China
[5]Energy Systems Division, Argonne National Laboratory, IL 60439, United States

Corresponding author: YU ZHANG (e-mail: zhangxiujade@163.com).

This work was supported by the Key Projects of the National Natural Science Foundation of China under Grant 51437003 and the Jilin science and technology development plan project of China under Grant 20160623004T, 20180201092GX.

**ABSTRACT** Because of the non-uniformity of the electric power CPS network and the dynamic nature of the risk propagation process, it is difficult to quantify the critical point of a cyber risk explosion. From the perspective of the dependency network, this paper proposes a method for quantitative evaluation of the risk propagation threshold of power CPS networks based on the percolation theory. First, the power CPS network is abstracted as a dual-layered network-directed unweighted graph according to topology correlation and coupling logic, and the asymmetrical balls-into-bins allocation method is used to establish a "one-to-many" and "partially coupled" non-uniform power CPS characterization model. Then, considering the directionality between the cyber layer and the physical layer link, the probability of percolation flow is introduced to establish the propagation dynamic equations for the internal coupling relationship of each layer. Finally, the risk propagation threshold is numerically quantified by defining the survival function of power CPS network nodes, and the validity of the proposed method is verified by the IEEE 30-bus system and 150-node Barabsi-Albert model.

**INDEX TERMS** Electric power CPS, interdependent network, Percolation probability, Propagation dynamics

## I. INTRODUCTION

With the advancement of smart grid strategy, a large number of electrical equipment, data collection equipment and computing equipment are interconnected through two physical networks: the power grid and the cyber network. Traditional power systems with physical equipment as the core have gradually evolved into highly coupled Cyber Physical Systems [1]. A power CPS integrates the physical environment of the computing system, communication network and power system through 3C technology to form a multi-dimensional and heterogeneous complex network system with real-time perception, dynamic control, resource optimization, cyber fusion and interdependence [2], [3]. It is because of this dependency that the security of the cyber system can significantly affect the operation of the physical system. The risks in the cyber system space may also lead to power outages in the power grid [4].

In recent years, experts and scholars at home and abroad have studied and summarized the causes and laws of power accidents over the years, and they have found that when the failure rate of cyber system components exceeds a certain level, power system accidents occur [5]. The existence of cyber system risks such as attack behavior, security risks, and risk explosion may lead to the abnormal operation or failure of components, and such failures may propagate from a single component to the entire power grid. Due to the high degree of coupling in a cyber-physical system, even if the risk is small, once it propagates, the butterfly effect it generates may propagate over a wide range, which will adversely affect the cyber system and the power grid. When the risk causes the loss of parts beyond a certain limit, it may even cause a large-scale blackout at a critical value.

Therefore, determining the critical condition of security risks that can widely propagate in the power CPS network or the assessment of a security risk propagation threshold is of important theoretical and practical significance. The security risk propagation threshold has always been a primary concern in the study of complex network theory [6]. In a typical complex network, there are two main methods for assessing security risk thresholds of a power CPS network:

1) One is the use of the dynamical equation of propagation based on the epidemiological propagation model [7], [8], wherein the SIS and SIR propagation models are the two most widely used classical propagation models [9], [10]. However, the use of this model requires the network under study to be a uniform single network, and the model has strict conditions for use and low universality. Although the actual power CPS network is a non-uniformly coupled network, the above method is no longer applicable.

2) Another method is to set up a time-domain discrete differential equation group for transmission of power based on the reduction theory [11], [12] and establish a time-domain discrete mathematical model for cyber flow using finite automata [13], [14]. In power CPS, there are essential differences in the transmission mechanism between power flow and cyber flow. It is difficult for this method to fully consider the characteristics of these two flows [15], and this method ignores the overall dynamics of the network. The analytical expressions are mostly implicit function expressions, and it is difficult to present the solving method.

Based on existing research, this paper considers the directionality and coupling relationship of the link lines between the cyber network and the physical network in the power CPS network and establishes the power CPS network characterization model. Then, on this basis, this paper uses the theory of percolation flow to propose the risk propagation dynamic model of a power CPS network and provide a numerically quantitative evaluation of the model by defining the survival function of the power CPS network node. Finally, this paper illustrates the effectiveness of the method by practical examples.

## II. CHARACTERIZATION MODEL OF NON-UNIFORM POWER CPS NETWORK

The quantitative assessment of the risk propagation threshold for power CPS networks first requires an effective and realistic network model. From the perspective of interdependent networks, there is an interdependence relationship between cyber networks and physical networks. That is, the cyber network is the "brain" and control system of the physical networks; the physical network provides energy for the cyber network [16], and they are interdependent and coupled into a two-tiered complex physical cyber fusion system. Most existing power CPS network modeling uses "one-to-one" coupling [17]; however, in an actual power system, a cyber node can only control one physical node, and one physical node can provide energy for multiple cyber nodes. In terms of the number of deployments and control methods, the number of cyber nodes is much larger than the number of physical nodes. Therefore, the dependency between the physical network and the cyber network node is "one-to-many" coupling. There are such nodes in the actual power system: it is highly autonomous and does not depend on the coupling network to be able to operate normally. Therefore, the "partial coupling" of nodes is more in line with the characteristics of the actual power CPS network.

Because there are many differences in the connection modes and types of equipment for power CPS, to construct the characterization model of a non-uniform power CPS, the following definitions and assumptions are made based on complex network theory in this paper:

1) Taking the plant station level as the research unit, the cyber network (including the cyber systems and dispatch centers of each power station) and physical sites (including power plants, substations, and converter stations) are considered to be equivalent cyber nodes and physical nodes, respectively.

2) The communication line between the cyber sites is equivalent to the edge of the cyber network. The transmission line between the physical sites is equivalent to the edge of the physical network.

3) Considering the directionality and dependencies between the links of the physical network and the cyber network, the links between the layers are undirected edges, and the edges between different layers are directed edges.

4) Loops and multiple edges on the line are merged.

Based on the above definitions and assumptions, the topology of the cyber network and the physical network is abstracted based on the complex network theory and expressed as two unweighted partial directed graphs $G_c$ and $G_p$, where $G_c$ represents the cyber network and $G_p$ represents the physical network.

### A. PHYSICAL LAYER CHARACTERIZATION

The physical layer model can be abstracted as a complex network unweighted graph, $G_p = <V_p, E_p>$, where $V_p$ is the node (power plants, substations, and converter stations), $E_p$ is the edge (transmission lines), $V_p = \{1,2,3....,N_p\}$ is the node set of the physical network, $E_p = \{E_{pij}\}$ is the set of connection edges of the physical network, and $A_p = (a_{pij})$ is the adjacency matrix of the physical network. Edges between physical layer nodes do not consider the direction, nor do they consider the capacity between the edges. In the coupling model, if a physical node fails, the cyber node that depends on its energy also fails.

## B. CYBER LAYER CHARACTERIZATION

The cyber network node is the control and processing center of the corresponding physical layer network node. In the cyber model, all relevant functions are considered to be completed in the abstract node. Similar to the physical layer model, the cyber layer model is abstracted as a complex network weightless graph, $G_c = <V_c, E_c>$, where $V_c$ is the node (server, computing device, and data acquisition device), $E_c$ is the edge (communication line), $V_c = \{1,2,3...,N_c\}$ is the node set of the cyber network, $E_c = \{E_{cij}\}$ is the set of connection edges of the cyber network, and $A_c = (a_{cij})$ is the adjacency matrix of the cyber network. Edges between cyber layer nodes also do not consider directions. In the coupling model, if a cyber node fails, the cyber node cannot communicate with its neighbor nodes. At the same time, because the invalid cyber node controls the corresponding physical node, its corresponding physical node also fails.

## C. CYBER-PHYSICAL COUPLING CHARACTERIZATION MODEL

Through the above modeling method, an independent physical layer model and cyber layer model are obtained. Because there are "one-to-many" and "partially coupled" dependencies between the physical network and the cyber network, it is necessary to couple the two interdependent networks into a two-layer network model using an effective method. A large amount of data shows that the physical network and cyber network are in line with the characteristics of scale-free networks, and the degree distribution of nodes meets the power-law distribution characteristics. This paper builds a two-layer coupling network based on the asymmetric Balls-into-Bins distribution algorithm [18]: only one node in $G_c$ supports linking to a node in $G_p$, and each node in $G_p$ can link to multiple $G_c$ nodes.

The sizes of the physical network and the cyber network are denoted by $|G_p|$ and $|G_c|$, respectively. To allocate links between the physical layer and the cyber layer nodes, it is assumed that the nodes in $G_p$ are all bins, the nodes of $G_c$ are balls, and $|G_c|$ is independent and evenly put in $|G_p|$. The probability that each ball is assigned to the $i$-th bin is $1/|G_p|$. For any $1 \le i \le |G_p|$, all balls are assigned to the $i$-th bin by $l_{pi}$. $|G_c|$ is independent and evenly put in $|G_p|$. The probability that each ball is assigned to the $i$-th bin is $1/|G_p|$. For any $1 \le i \le |G_p|$, all balls are assigned to the $i$-th bin by $l_{pi}$,

$$Pr(l_{p,i} = k) = \binom{|G_c|}{k} \cdot \left(\frac{1}{|G_p|}\right)^k \cdot \left(1 - \frac{1}{|G_p|}\right)^{|G_c|-k} \quad (1)$$

The $G_c$ nodes supported by each node in $G_p$ obey the binomial distribution B $(|G_c|, 1/|G_p|)$. In this system model, the link from $G_p$ to $G_c$ is directional. If there are two edges from $G_p$ to $G_c$, it indicates that there is interdependence between the two points in $G_p$ and $G_c$. For the ith node in $G_p$, one link is randomly selected from its $k$ links as a bidirectional link. This means that the $G_p$ node supports the $G_c$ node and the $G_c$ node controls the $i$-th $G_p$ node. For any $1 \le i \le |G_p|$, $1 \le j \le |G_c|$, and defined event $\varepsilon_{ij}$, the $j$ node in $G_c$ controls the $i$ node of $G_p$. For any defined event $\varepsilon_j$, select the $j$ node in $G_c$ as the operation center. When $\varepsilon_{i,j} \cap \varepsilon_{h,j} = \varnothing$ and $i \ne j$, then:

$$Pr(\varepsilon_j) = \sum_{i=1}^{|G_p|} Pr(\varepsilon_{i,j}) = \sum_{i=1}^{|G_p|} \sum_{k=1}^{|G_c|} Pr(l_{p,i} = k) \cdot \frac{k}{|G_c|} \cdot \frac{1}{K} \quad (2)$$

Because each node in $G_c$ has only one internal link and $G_c$ is a scale-free network, the node degree distribution in $G_c$ follows the Bernoulli distribution, and formula (2) is improved as:

$$Pr(\varepsilon_j) = \frac{|G_p|}{|G_c|} \cdot \sum_{i=1}^{|G_c|} Pr(l_{p,i} = k) = \frac{|G_p|}{|G_c|} \quad (3)$$

It can be seen that the probability of two-way links in the network is $|G_p|/|G_c|$. Through the above method, the characterization model for a "one-to-many" partial coupling power CPS non-uniform network can be established, as shown in Fig. 1.

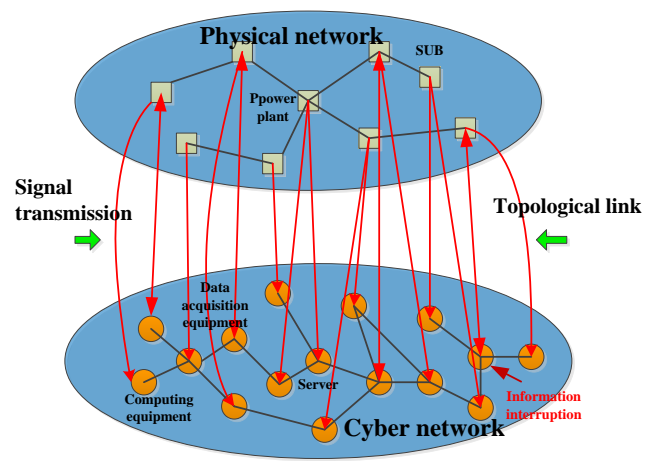

**FIGURE 1. Power CPS coupling model**

## III. CONSTRUCTION OF RISK COMMUNICATION DYNAMICS MODEL BASED ON PERCOLATION PROBABILITY

At present, there have been many simplifications for the network. It is believed that physical node failure causes the coupled cyber node to fail, and the failure of the cyber node will also cause the physical node to fail [19]. However, in a power cyber-physical system, important nodes (such as substations) widely use an uninterrupted power supply (such as UPS). Thus, node failure does not affect the normal operation of the power system for a short period of time [20]. Therefore, when considering the control and dependencies of the power CPS, when the cyber node fails, the communication line may not be disconnected, and the physical node failure may not necessarily cause the cyber node to fail. For this reason, this paper considers the directionality and dependence of the coupling between the cyber network and the physical network. If node $A$ is controlled by node $B$, node $A$ will fail when node $B$ fails; if node $A$ is linked to node $B$, node $A$ does not depend on node $B$. When node $B$ fails, node $A$ does not fail. Because the propagation of risk in the network is directional, hierarchical, and dynamic, when studying the risk propagation mechanism in the power CPS network, the risk propagation process can be equated to the deletion of nodes or edges in the coupling network. Here, we introduce the probability of percolation to simulate the failure probability of the nodes between the networks. The seepage probability is based on the analysis probability based on the various structures of the graph and its evolution process. It compares the risk propagation process to the process in which the points and edges in the network are infected with a certain probability. That is, failure of the cyber node leads to the failure of the physical node in the coupled network with a certain probability $\Phi$, which leads to the propagation of the next level of percolation. Taking Fig. 2 as an example to simulate the failure process of cyber network and physical network coupling.

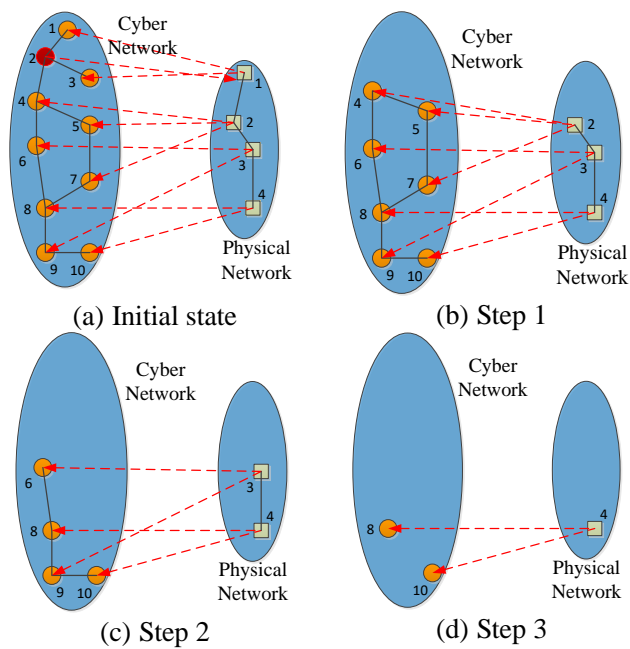

(a) Initial state      (b) Step 1

(c) Step 2      (d) Step 3

**FIGURE 2.** The process of interactive transmission between the cyber network and physical network

1) In the initial stage, there are 4 power nodes and 10 cyber nodes running in the coupling network. When node 2 is attacked in the cyber network, the failure propagates in the network due to the coupling relationship between the networks, as shown in Fig. 2.

2) In the largest connected subgraph in the cyber

network, node 2 in the cyber network is linked to node 1 in the physical network, and node 2 controls node 1. Therefore, when deleting cyber node 2, it is necessary to delete node 1 and the corresponding edges. Because node 1 in the physical network supplies power to nodes 1 and 3 in the cyber network at the same time, when the nodes in the physical network are deleted, nodes 1 and 3 in the cyber network must also be deleted, as shown in Fig. 2(b).

3) Judging the maximum connected subgraph in the physical network, it can be known that when the physical network node 2 fails, the remaining nodes are valid according to the dependence of the network, which further causes the nodes 4, 5, and 7 of the cyber network to be disabled. The physical network and the cyber network failed node are deleted, and the result is shown in Figure 2(c).

4) By analogy, node 3 in the physical network is deleted, and node 6 and node 9 in the cyber network are deleted at the same time, as shown in Fig. 2(d).

5) When the fault propagation stops, only node 8 and node 10 remain in the cyber network, and only node 4 in the physical network can still function normally.

Based on the percolation theory below, by mapping the intentional attacks of cyber nodes into random attacks, the propagation dynamics equations are established for the internal coupling relations of each layer. In this process, the fault propagation begins with $G_c$, then affects $G_p$, then returns to $G_c$, and so on, repeating the above process until the system is stable. Before performing the percolation operation, it is assumed that a node will perform the percolation operation only if the following conditions are met:

1) The node must be linked with a certain functional node, otherwise it is considered to be invalid, except for the autonomous node;

2) The node must belong to the largest connected subgraph of its own network; otherwise, it is considered to be invalid.

The communication process is shown in Fig. 3. The node randomly attacks a proportion of $G_c$ networks. Then, the number of remaining functional nodes in the $G_c$ network is $G_{c1}'=G_c \cdot (1-\Phi)=\mu_1' \cdot G_c$, and the number of nodes belonging to the largest connected subgraph in $G_{c1}'$ is set to $G_{c1}$, $G_{c1}=G_{c1}' \cdot F(\mu_1', \lambda_c)$. Among them, $\mu_1'$ and $\mu_1$ represent the remaining function nodes and the ratio of the largest connected group to all nodes, respectively. $F(\mu_1', \lambda_c)$ is the probability that a node belongs to the largest connected group and $\lambda_c$ is a power index.

According to the percolation flow, the propagation dynamics equations for each stage are established as shown in Fig. 3.

## A. $G_C$ NETWORK CLUSTER STATUS

First, a proportion $\Phi$ of nodes in $G_c$ are randomly removed to start the risk propagation of the power CPS network. At this stage, the $\Phi \cdot G_c$ nodes in $G_c$ are deleted and the internal links of these nodes are also deleted. When the link between nodes is deleted, $G_c$ is decomposed into clusters. At this time, the remaining number of functional nodes in the $G_c$ network is $G_{c1}'$.

$$| G_{c1}' | = | G_c | \cdot (1 - \Phi) = | G_c | \cdot \mu_1' \quad (4)$$

Because it is assumed that only the nodes contained in the largest connected subgraph can run, if this condition is not satisfied, some nodes and related edges are deleted. Here, we use $G_{c1}$ to represent the maximum connected subgraph after a $G_c$ fault:

$$| G_{c1} | = | G_{c1}' | \cdot F(\mu_1', \lambda_c) = | G_c | \cdot \mu_1 \quad (5)$$

In the formula, $\mu_1'$ and $\mu_1$ represent the remaining function nodes and the ratio of the largest connected group to all nodes. $F(\mu_1', \lambda_c)$ is the probability that a node belongs to the largest connected group, and $\lambda_c$ is a power index (same as below).

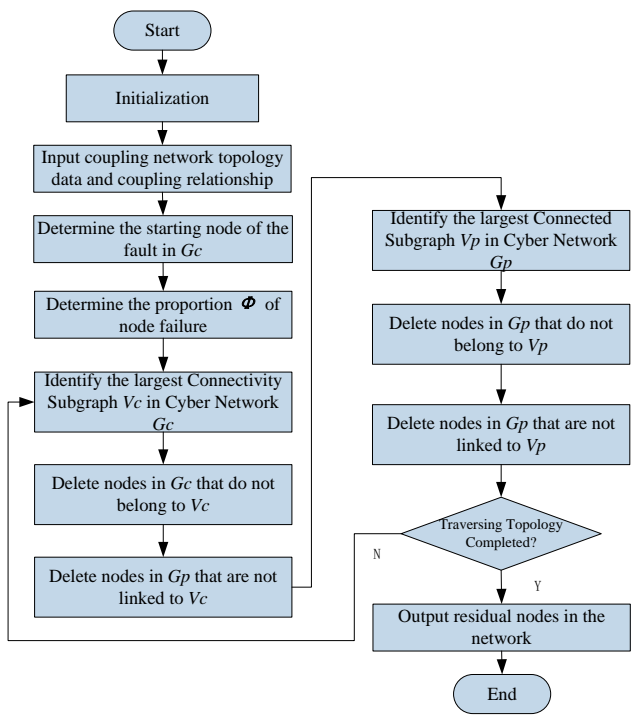

**FIGURE 3.** Flowchart of power CPS network penetration process

### B. $G_P$ NETWORK CLUSTER STATUS

In the first step, deleting the link between the nodes affects $G_p$. Because $G_c$ loses the link, some nodes and links in $G_p$ are also deleted. It is observed here that the links deleted in Phase 1 include unidirectional links and bidirectional links and now focus on bidirectional links because the nodes in $G_p$ depend on them. According to formula (3), the probability that each node in $G_p$ has a bidirectional link is $|G_p|/|G_c|$; therefore, the expected number of deleted bidirectional links is $(|G_c|-|G_{c1}|)$ ($|G_p|/|G_c|$), which is also the number of nodes that must be deleted in $G_p$. $G_{p2}'$ is used here to represent the remaining node set in $G_p$. There are still two-way links in the network:

$$| G_{p2}' | = | G_p | - (| G_c | - | G_{c1} |) \cdot \frac{| G_p |}{| G_c |} \quad (6)$$

$$= \mu_1' \cdot F(\mu_1', \lambda_c) \cdot | G_p |$$

$$\mu_2' = \mu_1' \cdot F(\mu_1', \lambda_c) \quad (7)$$

The largest connected group of $G_{p2}'$ is represented by $G_{p2}$.

$$| G_{p2} | = | G_{p2}' | \cdot F(\mu_2', \lambda_p)$$

$$= \mu_2' \cdot | G_p | \cdot F(\mu_2', \lambda_p) \quad (8)$$

$$\mu_2 = \mu_2' \cdot F(\mu_2', \lambda_p) \quad (9)$$

### C. DYNAMIC RECURSIVE EQUATION OF COUPLED NETWORK RISK PROPAGATION

Repeating the above process, the entire network will reach the final stable state, and a series of recursive equations can be used to represent the remaining components of the different stages of the network $G_c$ and $G_p$, as shown in Table I.

TABLE I
RECURSIVE EQUATIONS OF THE REMAINING COMPONENTS OF $G_c$ AND $G_p$ AT DIFFERENT STAGES

| Time phase | Cyber network $G_c$ | Physical network $G_p$ |
|---|---|---|
| Phase 1 | $\mu_1' = \Phi$ <br> $\mu_1 = \mu_1' \cdot F(\mu_1', \lambda_c)$ | |
| Phase 2 | | $\mu_2' = \mu_1' \cdot F(\mu_1', \lambda_c)$ <br> $\mu_2 = \mu_2' \cdot F(\mu_2', \lambda_p)$ |
| Phase 3 | $\mu_3' = \Phi \cdot \mu_2$ <br> $\mu_3 = \mu_3' \cdot F(\mu_3', \lambda_c)$ | |
| Phase 4 | | $\mu_4' = \mu_3' \cdot F(\mu_3', \lambda_c)$ <br> $\mu_4 = \mu_4' \cdot F(\mu_4', \lambda_p)$ |
| … | … | … |
| Phase 2j | | $\mu_{2j}' = \mu_1' \cdot F(\mu_{2j-1}', \lambda_c)$ <br> $\mu_{2j} = \mu_{2j}' \cdot F(\mu_{2j}', \lambda_p)$ |
| Phase 2j+1 | $\mu_{2j}' = \Phi \cdot \mu_{2j}$ <br> $\mu_{2j+1} = \mu_{2j+1}' \cdot F(\mu_{2j+1}', \lambda_c)$ | |

When the propagation behavior stops, the following equation is established:

$$\begin{cases} \mu_{2j+1}' = \mu_{2j+3}' = \mu_{2j-1}' \\ \mu_{2j}' = \mu_{2j+2}' = \mu_{2j-2}' \end{cases} \quad (10)$$

Because there are no more components in the largest connected subgraph in the two networks, here we order

$x = \mu_{2j+1}' = \mu_{2j+3}' = \mu_{2j-1}'$, $y = \mu_{2j}' = \mu_{2j+2}' = \mu_{2j-2}'$. At this point, we can obtain

$$\begin{cases} x = \Phi \cdot y \cdot F(y, \lambda_p) \\ y = \Phi \cdot y \cdot F(x, \lambda_c) \end{cases} \tag{11}$$

When $0 \le x$ and $y \le 1$ in both networks, the remaining ratio of the nodes in the final steady state can be calculated by

$$\begin{cases} \lim_{j \to \infty} \mu_{2j} = y \cdot F(y, \lambda_p) \\ \lim_{j \to \infty} \mu_{2j+1} = x \cdot F(x, \lambda_c) \end{cases} \tag{12}$$

At this point, a complete solution to the remaining components of $G_c$ and $G_p$ is obtained and the nontrivial solutions of x and y are solved. Then, the remaining number of nodes can be calculated.

## IV. QUANTITATIVE ESTIMATION METHOD FOR SURVIVAL FUNCTION-BASED PROPAGATION THRESHOLD

Through the analysis of the process of the risk propagation percolation, a series of recursive equations are obtained to solve the remaining components of the network. However, these recursive equations are implicit equations, and it is difficult to numerically quantify the risk propagation threshold of the network. This paper solves this problem by defining the survival function of the power CPS network node, the key part of which is the solution to $\mu_{2j}$ and $\mu_{2j+1}$, which is also equivalent to solving the x and y in the implicit function (11). Table II shows the symbols used in this section:

TABLE II
SYMBOLS USED

| Symbols | Description |
|---------|-------------|
| $d_p$ | The degree of an internal node of network P |
| $d_c$ | The degree of an internal node of network C |
| $G_{0,p}$ | Node degree distribution function in P |
| $G_{0,c}$ | Node degree distribution function in C |

The power CPS network threshold evaluation process is shown in Fig. 4.

First, we define the risk set of the power CPS coupling network $D_{cps} = < N_{cps}, E_{cps} >$, where $N_{cps}$ is a set of network nodes that are at risk, $E_{cps}$ is a set of directed edges. $N_{cps} = \{R, S\}$, and $N_{cps}$ is described by the node risk value $R$ and the node affected by the risk propagation factor S. $R \in [0,1]$ and the value of $S$ is represented by the probability that node $S_i$ fails. Risk sets can formally store and express the risk status of network nodes.

Then, the degree distribution function of the power CPS network $N$ is defined according to the distribution characteristics of degree-free network node degrees [21]:

$$G_{0,n}(u) = \sum_{k=0}^{\infty} pr(d_N = k) \cdot u^k \tag{13}$$

The $d_N$ follows the internal node degree distribution function of the network $N$, and $Pr(d_N = k)$ represents the probability that a node has $k$ internal links. Because the initially established network model is a scale-free network, the degree distribution of the network follows the power law distribution, that is, $Pr(d_N = k) = k \cdot k^{-\lambda}$, where $k$ is a constant and the power law index varies with different network structures.

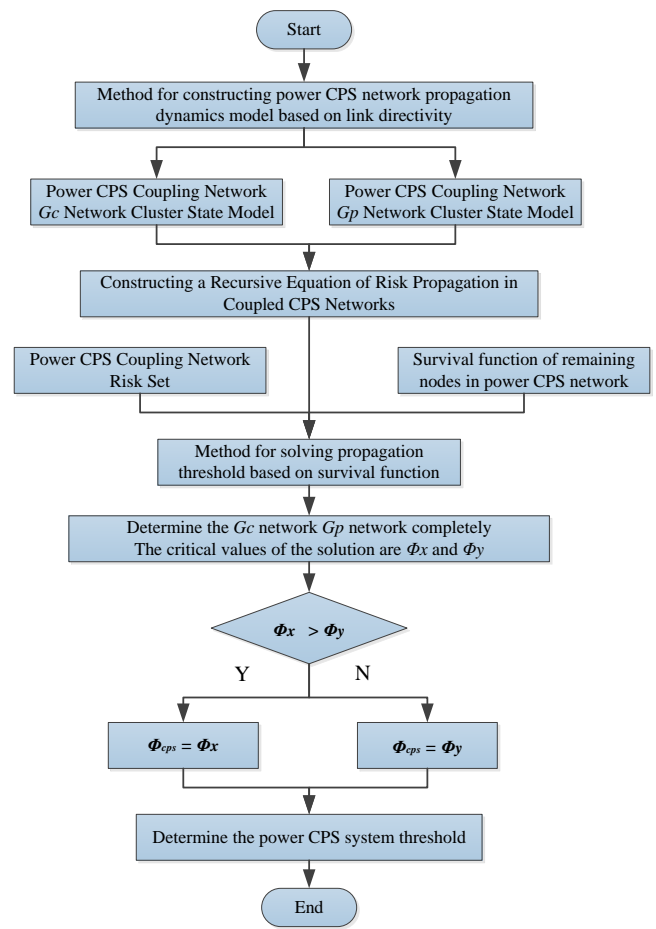

FIGURE 4. Flowchart of power CPS network threshold evaluation process

After removing the $1-\Phi$ proportional node from the power CPS coupling network $N$, the survival function of the remaining nodes is defined based on the risk set and the scale-free network degree distribution function and is represented by $F_N$.

$$\begin{cases} F_N(\Phi) = 1 - G_{0,N}(u) \\ u = 1 - \Phi + \Phi \cdot G_{0,N}(u) \end{cases} \tag{14}$$

wherein, $F_N(\Phi) \le 1$, for a single infinite scale-free network

$N$ that has a power index, when $2 < \lambda < 3$, $F_N(\Phi, \lambda) \in k \cdot \Phi^{1/(3-\lambda)}$, and $k$ is a constant.

Assume that the node degree distribution functions of $G_p$ and $G_c$ are $Pr(d_p = k) = k_a \cdot k^{-\lambda_p}$ and $Pr(d_c = k) = k_b \cdot k^{-\lambda_c}$, respectively, where $k_a$ and $k_b$ are constants. Combining the equations, we can obtain a set of equations:

$$\begin{cases} x = \Phi \cdot y \cdot F(y, \lambda_p) \\ y = \Phi \cdot y \cdot F(x, \lambda_c) \\ F(y, \lambda_p) = 1 - G_{0,p}(u_1) \\ u_1 = 1 - y + y \cdot G_{1,p}(u_1) \\ F(x, \lambda_c) = 1 - G_{0,c}(u_2) \\ u_2 = 1 - x + x \cdot G_{1,c}(u_2) \end{cases} \quad (15)$$

Suppose $F(y, \lambda_p) = k_1 \cdot y^{1/(3-\lambda_p)}$ and $F(x, \lambda_c) = k_2 \cdot y^{1/(3-\lambda_c)}$, where $k_1$ and $k_2$ are constants determined by the $G_p$ and $G_c$ network structures, respectively. Equation (15) can be simplified as follows:

$$\begin{cases} x = \Phi \cdot y \cdot k_1 \cdot y^{1/(3-\lambda_p)} \\ y = \Phi \cdot k_2 \cdot y^{1/(3-\lambda_c)} \end{cases} \quad (16)$$

After eliminating y, we can obtain

$$x = k_1 \cdot k_2^{(1+\frac{1}{3-\lambda_p})} \cdot \Phi^{2+(\frac{1}{3-\lambda_p})} \cdot x^{(1+\frac{1}{3-\lambda_p}) \cdot (\frac{1}{3-\lambda_p})} \quad (17)$$

The right side of (17) can be reduced to $C \cdot x^\eta$, where

$$\eta = (1 + \frac{1}{3-\lambda_c}) \cdot (\frac{1}{3-\lambda_p}) \quad (18)$$

When $2 < \lambda_p$, $\lambda_c < 3$, $\eta$ is far greater than 1. Here, we can see that (17) has a trivial solution $x = 0$, which shows that there is no node in the maximum connected sub-graph, that is, due to the risk of wide propagation in the network, the network is caused to completely collapse. Additionally, through computer simulation, it was found that when $x > 0$ and the removal ratio is greater than a certain value, the network starts to completely decompose. By setting $y = x$ as the reference line, if the implicit function curve crosses the reference line, it means the implicit function has a solution. As shown in Fig. 5, when $k_1 = 2$, $\lambda_c = \lambda_p = 2.5$, $\Phi = 0.2$, the curves converge at 0.635, which implies that under this network, the propagation threshold is 0.635, that is, when the removal ratio is greater than 0.635, the network is completely decoupled.

Therefore, when the following formula is satisfied, the threshold occurs:

$$k_1 \cdot k_2^{(1+\frac{1}{3-\lambda_p})} \cdot \Phi c^{2+(\frac{1}{3-\lambda_p})} = 1 \quad (19)$$

Thus, the critical value for the solution of $x$ is:

$$\Phi_x = \sqrt[2+\frac{1}{3-\lambda_p}]{k_1^{-1} \cdot k_2^{\frac{1}{\lambda_p - 3}}} \quad (20)$$

Similarly, we calculate the critical point of $y$:

$$\Phi_y = \sqrt[1+\frac{1}{3-\lambda_c}]{k_2^{-1} \cdot k_1^{\frac{1}{\lambda_c - 3}}} \quad (21)$$

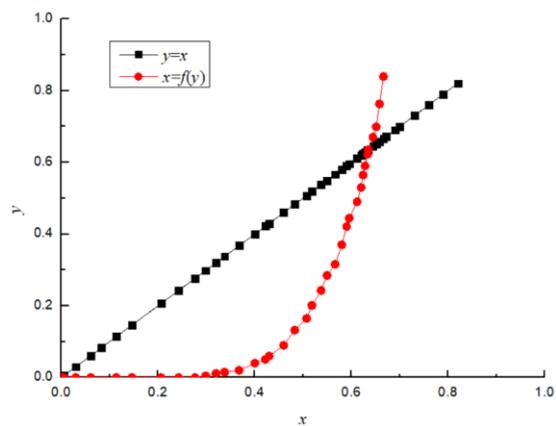

**FIGURE 5.** Effective solution of the curve

The condition $x$, $y \in [0,1]$ is satisfied, and the system threshold $\Phi_{cps}$ is the largest value in $[\Phi_x, \Phi_y]$. Therefore, (11) always has two solutions. 1) $x = y = 0$, which is a simple solution. 2) The other depends on $\Phi_{cps}$. If $\Phi > \Phi_{cps}$, the entire system will crash; otherwise, the system's largest connected sub-graph will continue to work.

## V. EXPERIMENTAL DESIGN AND ANALYSIS OF RESULTS

### A. EXPERIMENTAL DESIGN

The proposed propagation dynamics model is used to simulate the seepage process of non-uniform power CPS networks. The physical layer of the power cyber physical system is the IEEE 30 node standard model. The cyber layer is a 150-node scale-free network based on the Barabasi-Albert model with parameters $N = 150$, $m = 2$, $m_0 = 3$, and average degree $<k> \approx 4$. Asymmetric balls-into-bins allocation method is used between the two layers to establish a "one-to-many" coupling method. Based on this, a coupling network of 180-node power CPS constrained by various parameters is constructed. The coupling network has 30 power nodes, 107 communication nodes, 43 load nodes, 21 power lines, 125 information lines, and 84 coupling branches. Part of the network structure is shown in Figure 6. The association relationship is performed by using the Pajek simulation software with flat visualization, as shown in Fig. 7.

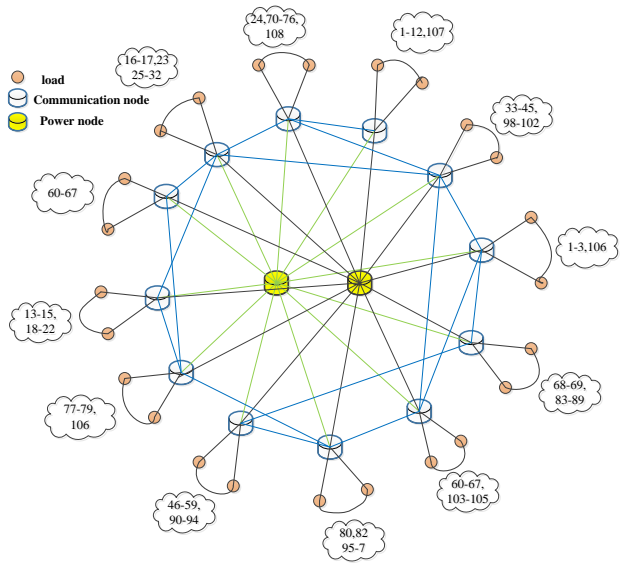

FIGURE 6. 180 node power CPS coupled network structure

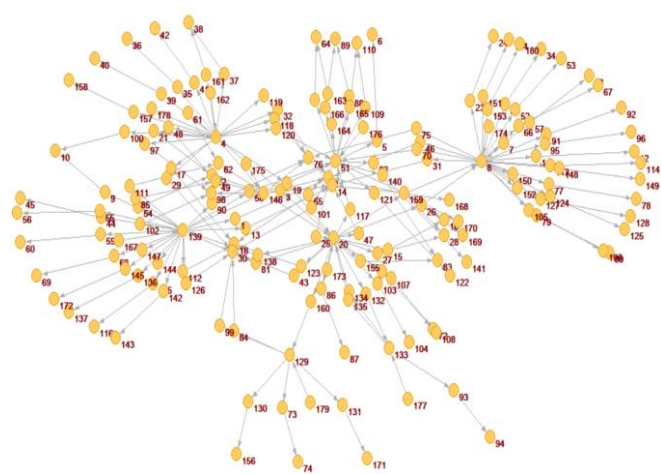

FIGURE 7. 180 node power CPS coupling topology

For the constructed 180-node power CPS "one-to-many" coupled network topology, the statistical topology degree distribution is shown in Table III.

TABLE III
DISTRIBUTION OF NETWORK NODE DEGREE

| Node degree | 1 | 2 | 3 | 4 | 5 | 6 | 7 | 10 | 12 | 20 | 22 | 24 | 25 | 28 |
|---|---|---|---|---|---|---|---|---|---|---|---|---|---|---|
| Number of nodes | 57 | 105 | 4 | 1 | 3 | 2 | 1 | 1 | 1 | 1 | 1 | 1 | 1 | 1 |

It can be seen from the table that most nodes in the network have degrees 1 and 2, and there are several nodes with large degrees in the network. Such nodes are important nodes in the network, and the nodes that are invalid or not functioning properly have a huge impact on the entire network.

### B. Example Analysis of The Evaluation Model

Random attacks and deliberate attacks were carried out on the established models. The number of remaining nodes in the power CPS network that accounted for the proportion of all nodes in different attack modes was completely de-listed from the entire network.

### 1) EVALUATION AND ANALYSIS OF NETWORK NODE FAILURE RATE UNDER RANDOM ATTACK

First, based on the constructed one-to-many coupling network topology of power CPS, the nodes in the cyber network are randomly removed, and the infiltration process in the entire coupled network is simulated by Java and Matlab to calculate the maximum connectivity in the network of each stage of seepage propagation. The number of subgraph nodes is the ratio of the original network nodes. As shown in Table IV, by continuously increasing the value of the removal ratio $\Phi$, it was found that when the $\Phi$ increased to 0.39, the network began to de-column; when $\Phi$ increased to 0.46, the maximum-connected sub-graph completely collapsed.

TABLE IV
THE VARIATION OF NODE FAILURE RATIO WITH $\Phi$

| Removal ratio $\Phi$ | 0.36 | 0.37 | 0.38 | 0.39 | 0.40 | 0.41 | 0.42 | 0.43 | 0.44 | 0.45 | 0.46 | 0.47 |
|---|---|---|---|---|---|---|---|---|---|---|---|---|
| Failure node ratio | 0.00 | 0.00 | 0.00 | 0.26 | 0.42 | 0.46 | 0.66 | 0.79 | 0.86 | 0.98 | 1 | 1 |

Through the analysis of the results, it can be found that there is a critical point $\Phi_{cps}$ in the power CPS coupling network. When the removal ratio $\Phi$ is less than the critical point, the proportion of the maximum connected subgraph failed nodes does not change; when the removal ratio $\Phi$ is greater than this critical point, the proportion of the failed nodes of the largest connected subgraph gradually increases, eventually equaling 1. Therefore, there is a threshold for the number of attack nodes in the power CPS network. Above this threshold, the network structure undergoes a qualitative change, all nodes in the cyber layer and the physical layer fail, and the fault range is extended to all nodes.

To further prove the above phenomenon and change the network parameter setting, let $G_c$=1000, $G_p$=10000, $m$=2, and $m_0$=3. Then, construct the BA-BA coupling network and calculate the power index, respectively: $\lambda_c = \lambda_p$ =2.2, $\lambda_c$ =2.2, $\lambda_p$ =2.33, $\lambda_c$ =2.2, $\lambda_p$ =2.5 in the three cases. The failure ratio of the $G_c$ network and the $\mu_{2j+1}$ phase of the $G_p$ network node changes with the removal ratio $\Phi$, where the values and simulation results are shown in Fig. 8.

The analysis results show that with the disengagement of the $G_c$ network, the coupled network $G_p$ is also decomposed, and the double-layer networks are, respectively, disjointed at the same stage. This also shows that in a dependent network, when a network is attacked, the network coupled with it is also affected by the same strength. In addition, by analyzing the network decomposition process of the $\mu_{2j}$ phase and the $\mu_{2j+1}$ phase, it can be found that when the failure rate of the single-layer network node exceeds the threshold, the entire coupling network is completely disjointed.

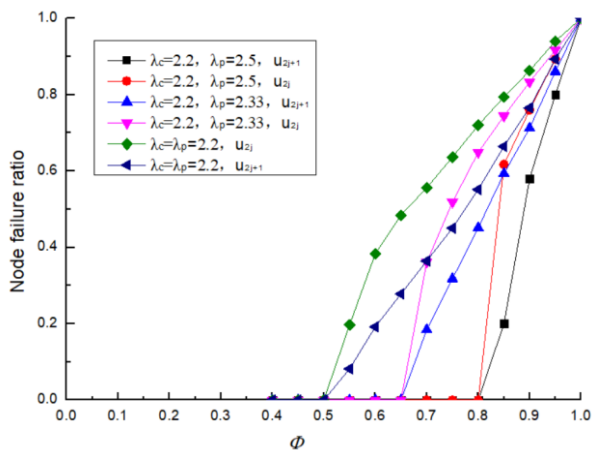

**FIGURE 8.** The variation of node failure ratio at each stage of $\Phi$

In the case where the above three network parameters are unchanged, the change of the failure ratio of the entire coupled network node with the removal ratio $\Phi$ is further analyzed. The simulation results are shown in Fig. 9.

Through single-layer network and coupled network simulation analysis, it can be concluded that because the nodes in the electric power CPS dependent network must rely on the nodes of other networks, when a network is attacked, the coupled network is also implicated, resulting in a cascading failure phenomenon. For different structures in the "one-to-many" coupled node network, there is still a network threshold phenomenon, and when the nodes of the same proportion are removed, the network failure ratios of different structures are also different. For networks with different structures, the network propagation thresholds are generally different.

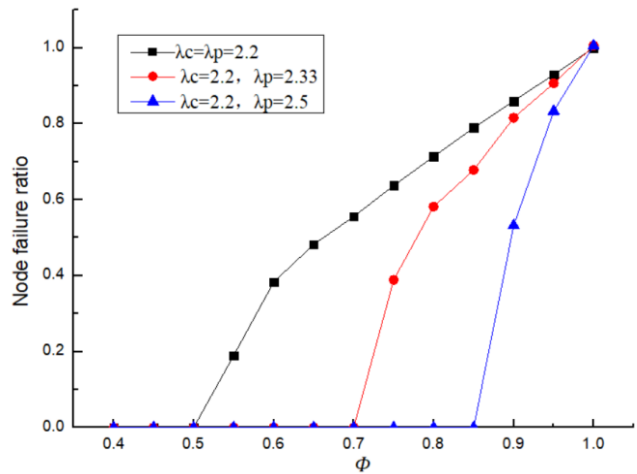

**FIGURE 9.** The variation of node failure ratio with $\Phi$

### 2) EVALUATION AND ANALYSIS OF NETWORK NODE FAILURE RATE UNDER DELIBERATE ATTACK

Similarly, a deliberate attack method was used for the above experiment to further analyze the threshold of the coupled network. It can be seen from Table III that there are 22, 24, 25, and 28 nodes with large degrees in the network. When the four nodes are deliberately removed, the network is quickly disconnected, and the entire network is immediately paralyzed. When an attacker has a deep understanding of the network topology, a deliberate attack on an important node in the network will have a destructive effect on the coupled network. When deliberately removing a small degree node, the network unwinding speed is equivalent to that when randomly removing a node. At this time, the change of the proportion of the remaining nodes of the maximum connectivity subgraph with the removal ratio $\Phi$ is shown in Table V. When $\Phi$ increases to 0.35, the network begins to de-collapse; when $\Phi$ increases to 0.41, the largest Unicom sub-graph completely collapses.

TABLE V
THE VARIATION OF NODE FAILURE RATIO WITH $\Phi$

| Removal ratio $\Phi$ | 0.33 | 0.34 | 0.35 | 0.36 | 0.37 | 0.38 | 0.39 | 0.40 | 0.41 | 0.42 | 0.43 | 0.44 |
|---|---|---|---|---|---|---|---|---|---|---|---|---|
| Failure node ratio | 0.00 | 0.00 | 0.38 | 0.53 | 0.62 | 0.77 | 0.84 | 0.96 | 1.00 | 1.00 | 1.00 | 1.00 |

Using the same network parameters, let $G_c$ =1000, $G_p$ = 10000, $m = 2$, and $m_0 = 3$ and construct $\lambda_c = \lambda_p = 2.2$, $\lambda_c = 2.2$, $\lambda_p = 2.33$, $\lambda_c = 2.2$, $\lambda_p = 2.5$ in the coupling scale-free network. In the deliberate attack mode, by deliberately removing the node of the $f_\Phi$ ratio, the coupling network $\mu_{2j}$ stage $G_p$ network and $\mu_{2j+1}$ stage $G_p$ network node failure ratio changes with the removal ratio $\Phi$ value. The simulation results are shown in Fig. 10. On this basis, we analyze the change of the node failure ratio with the removal ratio devaluation of the entire coupled network under the deliberate attack mode. The result is shown in Fig. 11.

Through the analysis of the simulation results, it can be found that in the deliberate attack mode, as the removal ratio increases slowly, the number of network failure nodes starts to increase slowly, but when $f_\Phi$ increases to the network threshold, the coupling network is quickly decomposed, and the growth rate becomes increasingly fast. When the network node removal ratio is only 0.2, the network is basically completely flawed. When further deliberately attacking a large number of nodes in the network, the entire network is immediately in a state of collapse, and the damage of the entire network is much larger than that of the random attack, and the fault propagation is more serious.

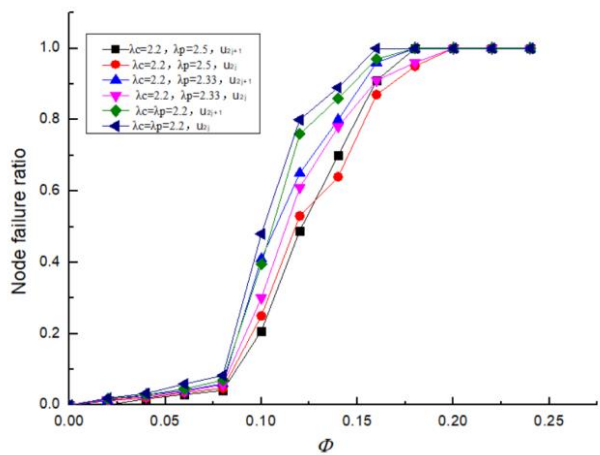

**FIGURE 10.** The variation of node failure ratio at each stage of $\Phi$

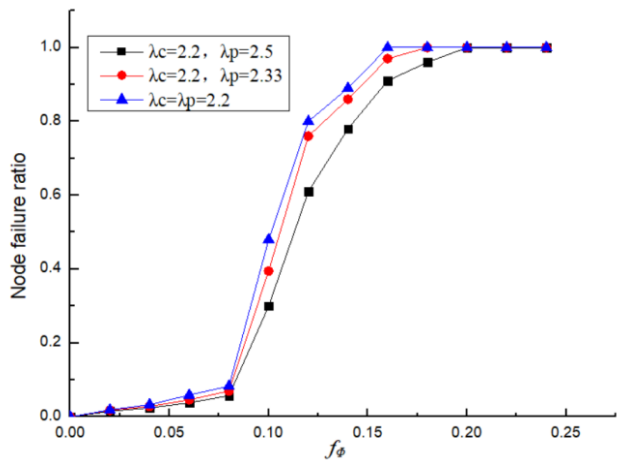

**FIGURE 11.** The variation of node failure ratio with $\Phi$

### 3) COMPARATIVE ANALYSIS OF NETWORK THRESHOLDS UNDER DIFFERENT COUPLING MODES

Under the above network parameter constraints, the network risk propagation thresholds under the random attack and deliberate attack mode of a "one-to-one" coupled network and "one-to-many" coupled network are compared. The simulation results are shown in Fig. 12.

The simulation results show that the model and method can effectively determine the risk propagation threshold of complex dependent networks. The network of different structures often has different propagation thresholds; compared to random attacks, the impact of deliberate attacks on the network is to a greater extent. Because the "one-to-one" coupling method oversimplifies the network structure, it is considered that when a node fails, the corresponding node also fails, and the directionality between the coupled network topologies is not considered, leading to a larger removal ratio. The network risk propagation threshold of the one-to-one coupling method is larger than that of the "one-to-many" network, and the network disjointing speed is too fast, whereas the "one-to-many" network considers the directionality of the coupled network, which is closer to the actual network situation. The resulting threshold can better and more accurately reflect the security performance of the network.

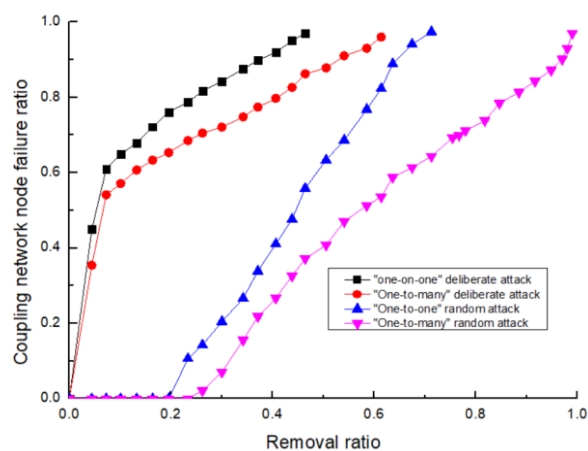

**FIGURE 12.** Network threshold comparison under different coupling modes

## VI. CONCLUSION AND FUTURE WORK

Based on the characteristics of power CPS in smart grids and the theory of interdependent networks, this paper takes into account the directionality and interdependence between the link between the physical network and the cyber network and proposes a characterization model for non-uniform power CPS network. Based on this model, this paper uses the theory of percolation to establish the dynamic model of the coupling of a power CPS and then proposes the survival function of the node to quantitatively evaluate this model. Simulation experiments show that the proposed method can effectively estimate the risk propagation thresholds of non-uniform and partially coupled networks. The threshold level is related to the scale of the power CPS network and the network topology (the power exponent and coupling mode of the network), and the risk propagation thresholds of power CPS networks with different structures are also different. The results of coupling network attacks are compared with random attacks and deliberate attacks. It is found that the intentional attack propagation behavior has a greater impact on the network. At the same time, the security risk propagation threshold can not only predict the critical point of risk explosion but also be used as a standard to measure network topology security. The greater the security risk propagation threshold of the network topology, the more difficult the risk spread is and the higher the security is of the network topology. According to the threshold constraint, the critical value of the security risk propagation burst of the power CPS can be defined, and the prediction of the security risk explosion under the complex system can be improved.

By further combining a new generation of artificial intelligence technology to consider the hardware characteristics and grid constraints of power cyber-physical systems, and it is the next step of research to reveal the risk propagation mechanism of power CPS network from the essence of mathematics, improve the predictive ability of network security risk outbreak, and formulate corresponding defense strategy according to the propagation path analysis.

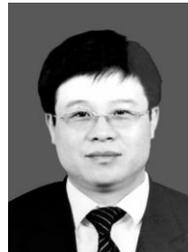

**ZHAOYANG QU** received his Ph.D degree in electrical engineering from China Northeast Electric Power University in 2006 and 2010, and his M.S degree in Dalian University of Technology in 1984 and 1988. He is currently a professor and doctoral tutorin the School of Information Engineering of Northeast Electric Power University. His interests include smart grid and power information processing, virtual reality and network technology.

He is a member of the China Electric Engineering Society Power Information Committee, the vice president of the Jilin Province Image and Graphics Society, the head of the Power Big Data Intelligent Processing Engineering Technology Research Center, and the Jilin Governor Baishan Scholar. He is also a top-notch innovative talent in Jilin Province and a young and middle-aged professional and technical talent with outstanding contributions. He presided over the completion of 2 national natural science funds, won the second prize of Jilin Province Science and Technology Progress Award, and wrote 46 papers on electric power information SCI/EI search.

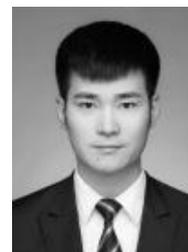

**YU ZHANG** was born in Henan, China. He received B.S degree in Computer science and technology from China Zhongyuan University of Technology in 2016. He studied for his M.S degree in NEEPU now. His interests include intelligent information processing, power cyber physical system and smart grid.

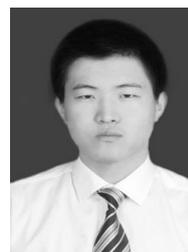

**NAN QU** was born in Jilin, China. He received his M.S. degree from Northeast Electrical Power University. He is an engineer in Nanjing Power Company now. His research interest is power system automation.

**IEEE** *Access*

Multidisciplinary ¦ Rapid Review ¦ Open Access Journal

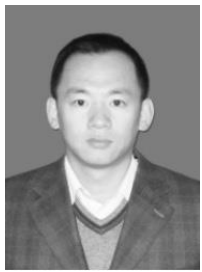

**LEI WANG** was born in Jilin, China. He is pursing his Doctor's degree of electrical engineering in Northeast Electrical Power University . He is an Associate professor at the School of Information Engineering. His research interest is information processing in Smart grid.

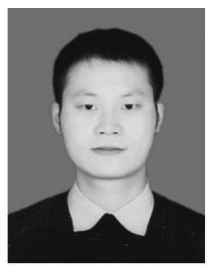

**YANG LI** (S'13–M'14–SM'18) was born in Nanyang, China. He received his Ph.D. degree in Electrical Engineering from North China Electric Power University (NCEPU), Beijing, China, in 2014. He is an Associate professor at the School of Electrical Engineering, Northeast Electric Power University, Jilin, China. Currently, he is also a China Scholarship Council (CSC)-funded postdoc with Argonne National Laboratory, Lemont, United States. His research interests include power system stability and control, integrated energy system, renewable energy integration, and smart grids. Dr. Li is the Associate Editor of IEEE Access.

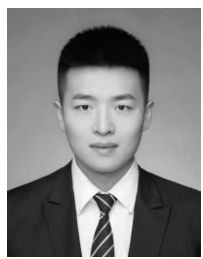

**YUNCHANG DONG** was born in Jilin, China. He received B.S degree from China Northeast Electric Power University (NEEPU) in 2012 and 2016. He studied for his M.S degree in NEEPU now. His interests include power cyber physical system and information processing of smart grid.